\documentclass[preprintnumbers,amsmath,amssymb,aps,prl]{revtex4}
\usepackage{amsmath}
\usepackage[mathcal]{eucal}
\usepackage{latexsym}
\usepackage{amssymb}
\newcommand*{\eps}{{\rlap{\lower2ex\hbox{$\,\,\tilde{}$}}{\epsilon_{ijk}}}}
\newcommand*{\EPS}{{\rlap{\lower2ex\hbox{$\,\,\tilde{}$}}{\epsilon_{i'j'k'}}}}
\newcommand*{\lmq}{{\rlap{\lower2ex\hbox{$\,\,\tilde{}$}}{\epsilon_{lmq}}}}
\newcommand*{\jmq}{{\rlap{\lower2ex\hbox{$\,\,\tilde{}$}}{\epsilon_{jmq}}}}
\newcommand*{\jql}{{\rlap{\lower2ex\hbox{$\,\,\tilde{}$}}{\epsilon_{jql}}}}
\newcommand*{\jlm}{{\rlap{\lower2ex\hbox{$\,\,\tilde{}$}}{\epsilon_{jlm}}}}
\newcommand*{\imq}{{\rlap{\lower2ex\hbox{$\,\,\tilde{}$}}{\epsilon_{imq}}}}
\newcommand*{\iql}{{\rlap{\lower2ex\hbox{$\,\,\tilde{}$}}{\epsilon_{iql}}}}
\newcommand*{\ilm}{{\rlap{\lower2ex\hbox{$\,\,\tilde{}$}}{\epsilon_{ilm}}}}
\newcommand*{\lmn}{{\rlap{\lower2ex\hbox{$\,\,\tilde{}$}}{\epsilon_{lmn}}}}
\newcommand*{\abc}{{\rlap{\lower2ex\hbox{$\,\,\tilde{}$}}{\epsilon_{abc}}}}
\newcommand*{\N}{{\rlap{\lower2ex\hbox{$\,\,\tilde{}$}}{N}}}
\newcommand{\tN}{{\rlap{\lower2ex\hbox{$\,\,\tilde{}$}}{N}}}
\newcommand*{\tM}{{\rlap{\lower2ex\hbox{$\,\,\tilde{}$}}{M}}}
\begin{document}
\title{Affine group representation formalism for four dimensional, Lorentzian, quantum gravity}

\author{Ching-Yi Chou}\email{l2897107@mail.ncku.edu.tw}
\address{Department of Physics, National Cheng Kung University, Taiwan}
\author{Eyo Ita}\email{ita@usna.edu}
\address{Department of Physics, US Naval Academy, Annapolis Maryland}
\author{Chopin Soo}\email{cpsoo@mail.ncku.edu.tw}
\address{Dept. of Physics, Nat. Cheng Kung U., Taiwan}
\input amssym.def
\input amssym.tex

%\address{$^{1}$\,\,Department of Physics, US Naval Academy,
%Annapolis, Maryland}
%\address{$^{2,3}$\,\,Department of Physics, National Cheng Kung University,
%Tainan 701, Taiwan}

%\ead{ita@usna.edu$^{1}$, cpsoo@mail.ncku.edu.tw$^{2}$,
%ccmaple014@gmail.com$^{3}$}

\bigskip

\begin{abstract}
Within the context of the Ashtekar variables, the Hamiltonian
constraint of four-dimensional pure General Relativity with
cosmological constant, $\Lambda$, is reexpressed as an affine
algebra with the commutator of the imaginary part of the
Chern-Simons functional, $Q$, and the positive-definite volume
element. This demonstrates that the affine algebra quantization
program of Klauder can indeed be applicable to the full Lorentzian
signature theory of quantum gravity with non-vanishing cosmological
constant; and it facilitates the construction of solutions to all of
the constraints. Unitary, irreducible representations of the affine
group exhibit a natural Hilbert space structure, and coherent states
and other physical states can be generated from a fiducial state. It
is also intriguing that formulation of the Hamiltonian constraint or
Wheeler-DeWitt equation as an affine algebra requires a
non-vanishing cosmological constant; and a fundamental uncertainty
relation of the form $\frac{\Delta{V}}{\langle{V}\rangle}\Delta
{Q}\geq 2\pi \Lambda L^2_{Planck}$ (wherein $V$ is the total volume)
may apply to all physical states of quantum gravity.
\end{abstract}
\maketitle

\section{Introduction}

Consistent quantization of gravity remains one of the most
challenging problems in theoretical physics.  To meet the standard
criteria for quantization via the canonical approach, a Hilbert
space for physical states, consistent with all of the constraints of
the theory, is needed. There is one known exact and explicit
solution to all of the constraints of pure General Relativity (GR)
with cosmological constant: the Chern-Simons state ${\psi}_{CS}$,
which as well exhibits a good semiclassical limit \cite{KOD,
POSLAMB, SOO1}. While lying in the simultaneous kernel of all of the
constraints for a particular operator ordering, this solution may
not meet the rigorous definition of a physical state owing to issues
of normalizability and unitarity raised by Witten and others
\cite{WITTEN}. In the present paper we shall reformulate the
Hamiltonian constraint of Lorentzian gravity, and in this context
demonstrate that physical states of the Hilbert space must belong to
a representation of an affine algebra. Some of the motivation for
this work comes from the results of \cite{SOO}, wherein one can
rewrite the Hamiltonian constraint completely in terms of
fundamental geometric entities which are the Chern-Simons functional
and volume element.

The main quantities of interest for the present paper will be the
imaginary part of the same Chern--Simons functional, $Q$, of the
Ashtekar connection, and the local and global volume operators
$V(x)$ and $V=\int V(x)$.  It will be demonstrated, within the
context of the Ashtekar variables, that the Hamiltonian constraint
for pure GR with cosmological constant can be reexpressed as an
affine algebra with the commutator of $Q$ and the volume element.
Another topic in quantum gravity which will be important for this
paper is the affine quantization program started by Klauder
\cite{KLAUDER}. In this approach, which was originally developed in
the metric representation, the spatial 3-metric $q_{ij}$ must
satisfy certain positivity requirements. As emphasized by Klauder,
these requirements are best implemented in the quantum theory via
the affine group representation wherein the solutions exhibit a
natural Hilbert space structure, and wherein coherent states and
other physical states can be generated from fiducial states.  The
present paper will demonstrate that the affine algebra quantization
program can indeed be applicable to the {\it full} theory of quantum
gravity and that {\it all} physical states must thus come from
representations of the affine algebra of the imaginary part of the
Chern-Simons functional and the positive-definite volume operator.
It is also intriguing that formulation of the Hamiltonian constraint
or Wheeler-Dewitt Equation as an affine algebra requires a
non-vanishing cosmological constant.

The organization of this paper is as follows:  In section 1 we
provide some background on the Ashtekar variables and their
quantization to set the stage.  Section 2 recapitulates some results
of \cite{SOO}, expressing the Hamiltonian constraint using Poisson
brackets involving the Chern--Simons functional $I_{CS}[A]$ and the
local volume element $V(x)$.  Section 3 introduces the relevant
basics of the affine group and its quantization, and sets up the
classical preliminaries.  Section 4 performs the actual quantization
using the affine group in conjunction with the relevant steps of the
algebraic quantization program due to Ashtekar \cite{ALGEBRAIC}, and
solutions of the Hamiltonian constraint as an affine algebra are
constructed from the affine group representations. Section 5 is a
brief summary of our main results. Further details and identities
related to affine group quantization are gathered in Appendix A. The
noteworthy identity that it is the imaginary part, rather than the
full Chern-Simons functional, that is really needed in the
reformulation of the Hamiltonian constraint as a affine algebra
(this ensures that the generators of the algebra are hermitian) is
relegated to Appendix B.

\section{The Ashtekar variables}

Let $M$ be a four dimensional spacetime manifold of topology
$M=\Sigma\times{R}$, where $\Sigma$ is a spatial 3-manifold of a
given topology embedded in $M$.  Then define on each $\Sigma$ the
canonical pair ($A^a_i,\widetilde{E}^i_a$), where $A^a_i$ is an
$SO(3)$ gauge potential, and $\widetilde{E}^i_a$ a densitized triad
of density weight 1 constructed from undensitized spatial triads
$e^a_i$\footnote{Index conventions are that symbols from the
beginning of the Latin alphabet $a,b,c,\dots$ label internal $SO(3)$
indices, while spatial indices are denoted by $i,j,k,\dots$.  For
internal indices, raised and lowered positions are equivalent, since
the group metric is the Euclidean 3-metric $\delta_{ab}$.  For
spatial indices, raised and lowered index positions are not
equivalent as the spatial metric is spatial 3-metric
$q_{ij}=e^a_ie_{aj}$.}.  These are given by
\begin{equation}
\label{THECONNECTION} A^a_i=\Gamma^a_i+\gamma{K}^a_i,\qquad
\widetilde{E}^i_a=\frac{1}{2}\widetilde{\epsilon}^{ijk}\epsilon_{abc}e^b_je^c_k;
\end{equation}
\noindent where $\gamma$ is the Barbero--Immirzi parameter,
$\Gamma^a_i$ is the spin connection compatible with $e^a_i$, and
$K^a_i$ is the triadic form of the extrinsic curvature of $\Sigma$.
Then the action for four dimensional gravity in the Ashtekar
variables can be written in 3+1 form as \cite{HOLST1, HOLST2,
HOLST3}
\begin{eqnarray}
\label{ACTION} I=\int{dt}\int_{\Sigma}d^3x\bigl(\frac{1}{\gamma
G}[\widetilde{E}^i_a\dot{A}^a_i+A^a_0G_a +N^iH_i] + N{\cal H}\bigr).
\end{eqnarray}
\noindent The equations of motion for the auxiliary fields
$A^a_0,N^i,N$ imply the vanishing of the Gauss' law, vector and
Hamiltonian constraints respectively $G_a$, $H_i$ and ${\cal H}$
given by
\begin{eqnarray*}
G_a&=&D_i\widetilde{E}^i_a=\partial_i\widetilde{E}^i_a+\epsilon_{ab}^cA^b_i\widetilde{E}^i_c;
\end{eqnarray*}
\begin{eqnarray*}
H_i&=&\widetilde{E}^j_aF^a_{ij}-\frac{1+\gamma^2}{\gamma}K^a_iG_a;
\end{eqnarray*}
\begin{eqnarray*}
{\cal
H}&=&\frac{1}{2G\sqrt{\vert\hbox{det}q\vert}}\widetilde{E}^i_a\widetilde{E}^j_b\Bigl(\epsilon^{ab}_cF^c_{ij}
+\frac{\Lambda}{3}\epsilon^{abc}\eps\widetilde{E}^k_c-2(1+\gamma^2)K^a_{[i}K^b_{j]}\Bigr)
\end{eqnarray*}
\begin{eqnarray}
\label{CONSTRAINTS}
+\frac{(1+\gamma^2)}{G\gamma^2}(\frac{\widetilde{E}^i_a}{\sqrt{\vert\hbox{det}q\vert}})\partial_iG^a.
\end{eqnarray}
\noindent Equation (\ref{ACTION}) provides the fundamental
nontrivial Poisson bracket\footnote{We have made the identification
$G\equiv{8}\pi{G_{\rm Newton}}/c^3$, to avoid carrying numerical
factors around.}
\begin{eqnarray}
\label{ACTION4}
\{A^a_i(x),\widetilde{E}^j_b(y)\}={\gamma}G\delta^a_b\delta^j_i\delta^{(3)}(x,y),
\end{eqnarray}
\noindent which in conjunction with the reality conditions
constitutes a basis for computation of the Hamiltonian dynamics of
four dimensional General Relativity and its quantization.\par
\indent For real $\gamma$ the connection $A^a_i$ is real and there
is no need to implement reality conditions in the Ashtekar-Barbero
formalism with generic real $\gamma$. On the other hand, the
Hamiltonian constraint in this case is difficult to implement on
account of the presence of extrinsic curvature squared and other
terms involving $\gamma$. For $\gamma=\pm{i}$ the aforementioned
extra terms vanish but the connection $A^a_i$ becomes complex. While
this yields a simple, polynomial Hamiltonian constraint, one must
impose reality conditions in order to get real General Relativity of
Lorentzian signature\footnote{We shall nevertheless show that the
generators of the affine algebra will still be hermitian}. From now
on, we will restrict consideration to $\gamma=i$ for concreteness,
and the (anti-)self-dual case will constitute the fundamental basis
for the results of this paper.

\subsection{Dirac Quantization procedure}

 There is no unique
prescription for the extrapolation from classical to quantum theory.
A standard approach to canonical quantization proceeds according to
the Heisenberg prescription, wherein one promotes the dynamical
variables to quantum operators ($A^a_i,\widetilde{E}^i_a)
\rightarrow (\widehat{A}^a_i,\widehat{\widetilde{E}^i_a}$) and
defines a set of unit vectors $\vert{\psi}\rangle\in{\bf H}_{Kin}$
on which the operators act.  These state vectors form a kinematic
Hilbert space ${\bf H}_{Kin}$, which is the Hilbert space at the
level prior to the implementation of any constraints.  All Poisson
brackets would become promoted to ${{2\pi} \over {ih}}$ times quantum
commutators, and so equation (\ref{ACTION4}) would become (for
$\gamma=i$) promoted to equal time commutation relations
\begin{eqnarray}
\label{ACTION13}
\bigl[\widehat{A}^a_i(x,t),\widehat{\widetilde{E}^j_b}(y,t)\bigr]=-(h{G}/2\pi)\delta^a_b\delta^j_i\delta^{(3)}(x,y).
\end{eqnarray}
\noindent The initial value constraints would become promoted to
operator constraints $\widehat{G}_a$, $\widehat{H}_i$ and
$\widehat{H}$ for a prescribed operator ordering.  According to the
Dirac procedure for constrained systems \cite{DIR}, the physical
states $\vert{\psi}\rangle\in{\bf H}_{Phys}$ are those elements of
${\bf H}_{Kin}$ which are annihilated by all of the quantum
constraints
\begin{eqnarray}
\label{PHYSICAL} \widehat{G}_a(x){|{\bf
\psi}\rangle}=\widehat{H}_i(x){|{\bf
\psi}\rangle}=\widehat{H}(x){|{\bf \psi}\rangle}=0.
\end{eqnarray}
\noindent This amounts to finding a set of gauge-invariant,
diffeomorphism-invariant functionals lying in the kernel of the
quantum Hamiltonian constraint, which admits a Hilbert space
structure.

In the present paper we will proceed via reformulation of the
Hamiltonian constraint and its consequent restriction on physical
states. This reveals the primacy of the Chern-Simons functional (or
rather its imaginary part) and the volume operators, and that they
form an affine algebra which must be satisfied by {\it all}
solutions of the local Hamiltonian constraint $H(x)=0$ or Wheeler
DeWitt equation. We will then quantize this algebra and construct an
associated Hilbert space respecting the constraints.

\section{Reformulation of the Hamiltonian constraint}

From the spatial curvature $F^a_{ij}$ let us define the Ashtekar
magnetic field of the connection $A^a_i$, an object of density
weight one given by
\begin{eqnarray}
\label{MAGNETICFIELD} \widetilde{B}^{ia}={1 \over
2}\widetilde{\epsilon}^{ijk}F^a_{jk}.
\end{eqnarray}
\noindent When the densitized lapse function
$\underline{N}=N/\sqrt{\vert\hbox{det}q\vert}$ is treated as a
fundamental auxiliary field, then the Hamiltonian constraint is a
polynomial constraint of density weight two, given by (we omit the
double-tilde on $H$ for simplicity; it should be clear from the
context)
\begin{equation}
\label{ACTION7} \frac{\delta{I}}{\delta\underline{N}(x)}= 0
\Leftrightarrow
H(x)=\eps\epsilon^{abc}\widetilde{E}^i_a\widetilde{E}^j_b\widetilde{B}^k_c
+{\Lambda \over
3}\eps\epsilon^{abc}\widetilde{E}^i_a\widetilde{E}^j_b\widetilde{E}^k_c=0.
\end{equation}
\noindent One of the steps in \cite{SOO} in reformulation of the
Hamiltonian constraint is that the Chern--Simons functional
$I_{CS}[A]$ of the connection $A^a_i$  has the Poisson bracket
\begin{equation}
\label{ACTION6} \{I_{CS}[A],\widetilde{E}^k_c(x)\}
=\int_{\Sigma}d^3y\widetilde{B}^j_b(y)\{A^b_j(y),\widetilde{E}^k_c(x)\}=iG\widetilde{B}^k_c(x).
\end{equation}
\noindent The curvature term of (\ref{ACTION7}) can thus be written
using this Poisson bracket by contracting (\ref{ACTION6}) with two
factors of $\widetilde{E}^i_a$ in antisymmetric combination and
using the definition of determinants, yielding
\begin{eqnarray}
\label{ACTION8} \{I_{CS}[A],\hbox{det}\widetilde{E}(x)\}&=&
{1 \over 2}\eps\epsilon^{abc}\widetilde{E}^i_a(x)\widetilde{E}^j_b(x)\{I_{CS}[A],\widetilde{E}^k_c(x)\}\nonumber\\
&=&{{iG} \over
2}\eps\epsilon^{abc}\widetilde{E}^i_a(x)\widetilde{E}^j_b(x)\widetilde{B}^k_c(x).\nonumber
\end{eqnarray}
\noindent Let us define a local volume squared operator $V^2(x)$,
given by
\begin{eqnarray}
\label{VOLUMESQUARED} V^2(x)={1 \over
6}\eps\epsilon^{abc}\widetilde{E}^i_a(x)\widetilde{E}^j_b(x)\widetilde{E}^k_c(x)
=\hbox{det}\widetilde{E}(x).
\end{eqnarray}
Then using (\ref{ACTION8}), equation (\ref{ACTION7}) can be written
in the following form
\begin{eqnarray}
\label{ACTION9}
\eps\epsilon^{abc}\widetilde{E}^i_a(x)\widetilde{E}^j_b(x)\widetilde{B}^k_c(x)+2\Lambda\hbox{det}\widetilde{E}(x)
=-\frac{2i}{G}\{I_{CS}[A],V^2(x)\} +2\Lambda{V}^2(x)=0.
\end{eqnarray}
\noindent Equation (\ref{ACTION9}) is a polynomial constraint, a
highly desirable feature from the standpoint of the quantum theory.
Indeed, the quantum Wheeler-Dewitt equation defined from the
fundamental commutation relations (\ref{ACTION13}) will have
symmetric factor-ordering, (omitting the hats for simplicity)
\cite{SOO}
\begin{eqnarray}
\label{SYMMETRICORDERING} -\frac{2}{\hbar G}[I_{CS}[A],V^2(x)] +
2\Lambda V^2(x) &=&
\frac{1}{3}\eps\epsilon_{abc}\bigl(\widetilde{E}^{ia}\widetilde{E}^{jb}\widetilde{B}^{kc}
+\widetilde{E}^{ia}\widetilde{B}^{jb}\widetilde{E}^{kc}+\widetilde{B}^{ia}\widetilde{E}^{jb}\widetilde{E}^{kc}\bigr)\nonumber\\
&&+ {\Lambda \over
3}\eps\epsilon^{abc}\widetilde{E}^i_a\widetilde{E}^j_b\widetilde{E}^k_c\nonumber\\
 &=:& H=0.
\end{eqnarray}
\noindent Equation (\ref{SYMMETRICORDERING}) is also the Weyl
ordering of $\widetilde{E}$ and $\widetilde{B}$.  Moreover, the
constraint algebra for a symmetric ordering of the constraints can
be shown to close formally, which is necessary for consistency of
Dirac quantization. Thus this form of the Hamiltonian constraint has
some nice properties. It has been pointed out in \cite{REGULATING}
that background independent field theories are ultraviolet
self-regulating if the constraint weight is equal to one but not for
other density weights.  This suggests that a more appropriate form
for the Hamiltonian constraint for quantization is as a density
weight one constraint.  A density weight one Hamiltonian constraint
can be accomplished by treating the lapse $N$ rather than the
densitized lapse $\underline{N}$ as the basic auxiliary field in the
action (\ref{ACTION}).  So vis-a-vis (\ref{ACTION7}), variation of
the lapse $N(x)$ yields
\begin{eqnarray*}
[\hbox{det}\widetilde{E}(x)]^{-\frac{1}{2}}
\bigl(\eps\epsilon^{abc}\widetilde{E}^i_a\widetilde{E}^j_b\widetilde{B}^k_c
+{\Lambda \over
3}\eps\epsilon^{abc}\widetilde{E}^i_a\widetilde{E}^j_b\widetilde{E}^k_c\bigr)
\end{eqnarray*}
\begin{eqnarray}
\label{SOTHAT}
=-\frac{2i}{G}[\hbox{det}\widetilde{E}(x)]^{-\frac{1}{2}}\bigl(\{I_{CS}[A],\hbox{det}\widetilde{E}(x)\}+iG\Lambda\hbox{det}\widetilde{E}(x)\bigr)=0.
\end{eqnarray}
\noindent For reasons related to diffeomorphism invariance which
will come up later, we will need to write the Hamiltonian constraint
in terms of the square root of $\hbox{det}\widetilde{E}(x)$.
% From the chain rule, the Poisson bracket
%\begin{equation}
%\label{SOTHAT1}
%\{I_{CS}[A],\sqrt{\vert\hbox{det}\widetilde{E}(x)}\vert\} ={1 \over
%2}(\vert\hbox{det}\widetilde{E}(x)\vert)^{-1/2}\{I_{CS}[A],\hbox{det}\widetilde{E}(x)\}
%\end{equation}
%\noindent holds, and substitution of (\ref{SOTHAT1}) into
Subject to non-vanishing $\det{\tilde E}$, (\ref{SOTHAT}) yields the
weight one Hamiltonian constraint as
\begin{eqnarray}
\label{ROOTCONSTRAINT}
\{I_{CS}[A],\sqrt{\vert\hbox{det}\widetilde{E}(x)}\vert\}=-{{iG\Lambda}
\over 2}\sqrt{\vert\hbox{det}\widetilde{E}(x)\vert}.
\end{eqnarray}
\noindent Equation (\ref{ROOTCONSTRAINT}) is however non-polynomial
in terms of the fundamental variable $\widetilde{E}^i_a$ on account
of the presence of the square root, which is an undesirable feature
from the standpoint of quantization; but it is polynomial in the
local volume operator
\begin{eqnarray}
\label{LOCALVOLUME} V(x)=\sqrt{\Bigl\vert{1 \over
6}\eps\epsilon^{abc}\widetilde{E}^i_a(x)\widetilde{E}^j_b(x)\widetilde{E}^k_c(x)\Bigr\vert},
\end{eqnarray}
\noindent the absolute value sign put in so that $V(x)$ is real both
for positive and negative triad orientations (while $V(x)$ is of
density weight one, we will omit the tilde symbol over it for
notational simplicity.  This should not lead to any confusion, and
the proper density weight should be understood from the context). In
terms of $V(x)$, the preferred local density weight one Hamiltonian
constraint can now be written (in $V$ and $A$) as
\begin{eqnarray}
\label{SOTHAT2} \{I_{CS}[A],V(x)\}=-{{iG\Lambda} \over
2}V(x)~\forall{x}.
\end{eqnarray}

\section{Affine group and quantum gravity with Ashtekar variables}

The importance of the affine group for quantum gravity was first
pointed out by Klauder in \cite{KLAUDER, KLAUDER1, KLAUDER2}; and
the general concepts of {\it continuous} representation theory for
the affine group have been developed by Klauder and Aslaksen in
\cite{ASLAKSEN}.  It is well known, from the Stone--Von Neuman
theorem, that there is only one irreducible representation up to
unitary equivalence of canonical, self-adjoint operators $p$ and $q$
satisfying the Weyl form of the canonical commutation relations
$[q,p]=-i$.  This representation, equivalent to the Schr\"odinger
representation, implies that the spectrum of both $p$ and $q$ cover
the whole real line.  The affine commutation relation for a single
degree of freedom takes the form \cite{ASLAKSEN}
\begin{eqnarray}
\label{AFFINERELATION} [D,q]=-iq,
\end{eqnarray}
\noindent where $D=(pq+qp)/2$ denotes the dilation operator and $q$
is the operator being dilated.  It has been established that there
exist two and only two unitarily inequivalent, irreducible
representations $\pi$ of the affine group $G_{aff}$, one
representation $\pi^{+}$ for which the spectrum of $q$ is positive
and another $\pi^{-}$ for which it is negative.  This provided the
motivation for a multidimensional generalization of the affine group
provided by Klauder, where the dilated objects become replaced with
operators corresponding to the spatial 3-metric $q_{ij}$.  As noted
in \cite{KLAUDER}, the metric must satisfy certain positivity
requirements which must be reflected in the quantum theory. Appendix
A is a supplement on the measure, inner product, and coherent states
associated with the affine group.
\par
\indent A comparison of (\ref{SOTHAT2}) with (\ref{AFFINERELATION})
reveals that the local Hamiltonian constraint (\ref{SOTHAT2}) is
nothing other than the Lie algebra of affine transformations of the
straight line.  More precisely, it is an infinite number affine Lie
algebras, one Lie algebra $g_{aff}(x)$ per spatial point $x$.  The
Chern--Simons functional $I_{CS}[A]$ plays the role of the dilator,
and $V(x)$ plays the role of the object being dilated.  By applying
the affine quantum gravity concept to the Ashtekar variables, we
will ultimately construct physical Hilbert spaces based upon the
representation $\pi^{+}$, thereby endowing the local volume operator
for $V(x)$ with a {\it positive spectrum}.
% (The treatment of both $\pi^{+}$ and $\pi^{-}$ would be tantamount to considering the two
%different orientations of the densitized triad $\widetilde{E}^i_a$.
%This allows the possibility of topology change in Lorentzian quantum
%gravity, where the quantum theories in each topological sector would
%be inequivalent.  By restricting attention to $V(x)\neq{0}$ we
%eliminate this possibility).
There is an important caveat though: $I_{CS}$ is not Hermitian due
to complex (anti-)self-dual Ashtekar variables; but we shall show
that in  (\ref{SOTHAT2}) only the imaginary part of the Chern-Simons
functional, $Q$, is relevant (the proof is relegated to Appendix B),
and thus {\it  the affine representation correspondence can be made
exact}.

We have shown the local density weight one Hamiltonian constraint
$H(x)=0$, as shown in (\ref{SOTHAT2}) can classically be written as
a Poisson bracket
\begin{eqnarray}
\label{KODAMA}
\{-iI_{CS}[A],V(x)\}=-\Bigl(\frac{G\Lambda}{2}\Bigr)V(x)~\forall~x.
\end{eqnarray}
\noindent
%For the purposes of the quantum theory, we will regard $V(x)$ to be a fundamental variable in lieu of the Ashtekar densitized triad $\widetilde{E}^i_a$, just as well as we will regard %the imaginary part of the Chern--Simons functional $Q=Im[I_{CS}[A]]$ to be a fundamental variable in lieu of the Ashtekar connection $A^a_i$.\footnote{This is analogous to the %holonomy--flux algebra (\ref{WILSON1}), which quantizes holonomies as fundamental quantities in lieu of the connection $A^a_i$.}\par
With the following definitions for the real and the imaginary parts
of the Chern--Simons functional,
\begin{eqnarray}
\label{KODAMA2} Y=Re[I_{CS}[A]];~~Q=Im[I_{CS}[A]],
\end{eqnarray}
\noindent it follows that at the classical level, the Hamiltonian
constraint (\ref{KODAMA}) is
\begin{eqnarray}
\label{KODAMA4} -i\{Y,V(x)\}+\{Q,V(x)\}=-\Bigl({{G\Lambda} \over
2}\Bigl)V(x).
\end{eqnarray}
\noindent We will now make use of a remarkable identity
\begin{eqnarray}
\label{REMARKABLE} \{-iI_{CS}[A],V(x)\}=\{Im[I_{CS}[A]],{V}(x)\},
\end{eqnarray}
\noindent namely that the Poisson bracket of $V(x)$ with the
Chern--Simons functional $I_{CS}[A]$ is the same as the Poisson
bracket of $V(x)$ with its imaginary part. This is proven in
Appendix B.  This means that the first term on the left hand side of
(\ref{KODAMA4}) is zero, which implies the following fundamental
Poisson brackets
\begin{eqnarray}
\label{KODAMA5} \{Q,V(x)\}=-\Bigl({{G\Lambda} \over
2}\Bigr)V(x);~~\{Y,V(x)\}=0.
\end{eqnarray}
\noindent Thus the imaginary part $Q$ of the Chern--Simons
functional $I_{CS}$ admits an affine Lie algebraic structure with
$V(x)$ for each $x$, while the real part $Y$ commutes with
$V(x)$.\par
%We will regard the latter as proportional to the identity operator in the quantum theory.\par
%\indent We will now return to a point which we postponed from
%(\ref{SOTHAT2}) vis-a-vis (\ref{ACTION9}).
Noting that $Q = \frac{1}{2i} (I_{CS} - I^\dagger_{CS})= Q^\dagger$,
and $V^\dagger = V$; it follows from Eq.(\ref{SYMMETRICORDERING})
that  (in the quantum operators we omit the hats for simplicity,
they will be restored when necessary)
 \begin{eqnarray}
\label{RIGOROUS}
 [Q,V^2(x)]= -2i\lambda V^2(x); \qquad \lambda=\frac{h{G}\Lambda}{4\pi}.\end{eqnarray}
In order to formulate the  Hamiltonian constraint as a weight one
scalar density and to make contact with the affine representation,
we postulate
\begin{eqnarray}
\label{RIGOROUS1} [Q,V(x)]=-{i\lambda}V(x)
\end{eqnarray}
as the equation of the Hamiltonian constraint defining physical states. It holds at classical Poisson bracket level, and it is also a sufficient condition %(albeit not necessary)
for (\ref{SYMMETRICORDERING}) and (\ref{RIGOROUS}).

\section{Algebraic quantization}

We shall now proceed with the quantization of the theory, bringing
in the relevant steps from the algebraic quantization program
\cite{ALGEBRAIC},\cite{ASHTEKAR}.\par \noindent (i) Step 1: Our
first step will be to identify an appropriate set $S$ of $SO(3)$
gauge-invariant classical objects, which is closed under Poisson
brackets and under complex conjugation
\begin{eqnarray}
\label{KODAMA6} S=\{I,Q,V(x)\}_{\forall{x}\in\Sigma}.
\end{eqnarray}
\noindent Encoded in the requirement that the set $S$ be closed
under Poisson brackets is the solution to the local Hamiltonian
constraint (\ref{RIGOROUS1}).  Combined with the fact that $S$ is
invariant under the identity-connected component of complex $SO(3)$
transformations, this will address the Gauss' law and Hamiltonian
constraints.  We will address the diffeomorphism constraint
shortly.\par \noindent (ii) Step 2: Each function in $S$ will be
regarded as an elementary classical variable which is to have an
unambiguous quantum analogue.  Since $Q$ and $V(x)$ are composite
objects constructed respectively from purely {\it commuting}
coordinate $A^a_i(x)$ and momentum $\widetilde{E}^i_a(x)$ variables
from the original Ashtekar phase space, then their quantum analogues
will be {\it free of ordering ambiguities}.\par \indent With each
element $I,Q,V(x)\in{S}$ we associate an abstract operator
$\widehat{I},\widehat{Q},\widehat{V}(x)$, which defines the free
associative algebra $B_{aux}$ generated by these elementary quantum
operators.  Upon this we impose the commutation relation, consistent
with the Heisenberg--Dirac promotion of classical Poisson brackets
to quantum commutators
\begin{eqnarray}
\label{KODAMA7} \{Q,V(x)\}=-\Bigl({{G\Lambda} \over
2}\Bigr)V(x)\longrightarrow{1 \over
{(ih/2\pi)}}[\widehat{Q},\widehat{V}(x)] =-\Bigl({{G\Lambda} \over
2}\Bigr)\widehat{V}(x).
\end{eqnarray}
\noindent Then using the definitions (\ref{KODAMA2}), the quantum
Hamiltonian constraint can be written as an infinite number of
affine commutation relations, one affine commutation relation per
spatial point $x$\footnote{For precise correspondence to the
standard form of the affine algebra (\ref{AFFINERELATION}), we may
divide both sides of (\ref{RIGOROUS1}) by $\lambda$, rescaling
$Q\rightarrow{Q}/\lambda$ in units of the dimensionless quantity
$\lambda=h{G}\Lambda/4\pi$. }
\begin{eqnarray}
\label{KODAMA8}
[\widehat{Q},\widehat{V}(x)]=-i\lambda\widehat{V}(x);~~[\widehat{Q},\widehat{Q}]=[\widehat{V}(x),\widehat{V}(y)]=0.
\end{eqnarray}
\noindent From the commutation relations (\ref{KODAMA8}) can be
immediately written down the following exponentiated point-wise
relations for any real numbers $a$ and $b$
\begin{eqnarray}
\label{KODAMA101}
e^{-ia\widehat{Q}}\widehat{V}(x)e^{ia\widehat{Q}}=e^{-\lambda{a}}\widehat{V}(x);~~
e^{-ib\widehat{V}(x)}\widehat{Q}e^{ib\widehat{V}(x)}=\widehat{Q}+\lambda{b}\widehat{V}(x).
\end{eqnarray}
\noindent Since $V(x)$ is a locally defined object of density weight
one, it is not diffeomorphism invariant.  It transforms under
spatial diffeomorphisms parametrized by any smooth vector field
$\xi^i\in{C}^{\infty}(\Sigma)$ as
\begin{eqnarray}
\label{KODAMA11}
\delta_{\vec{\xi}}V(x)={\pounds}_{\vec{\xi}}V(x)=\partial_i(\xi^i(x)V(x))\neq{0},
\end{eqnarray}
\noindent which means that the local Hamiltonian constraint $H(x)=0$
is also not diffeomorphism invariant and is of density weight 1. But
we want our states to be solutions of the diffeomorphism constraint
$H_i(x)=0$ in addition to the Gauss' law constraint $G_a(x)=0$,
while at the same time lying in the kernel of $H(x)$. That is, we
want {\it physical} states, or elements of the physical Hilbert
space ${\bf H}_{Phys}$. For diffeomorphism-invariant statements, it
will be apposite to construct a {\it global} total volume functional
$V$ from the {\it local} function $V(x)$, given by
\begin{eqnarray}
\label{KODAMA12} V=\int_{\Sigma}d^3xV(x).
\end{eqnarray}
\noindent Note that $V$, the volume of 3-space $\Sigma$, is a
diffeomorphism-invariant quantity, and so is $Q$.

On integrating (\ref{KODAMA8}) over $\Sigma$, then
$[\widehat{Q},\widehat{V}]=-i\lambda\widehat{V}$, the global form of
affine algebra involving $Q$ and the total volume $V$ must also
hold. So the following exponentiated relations, a weaker form of
(\ref{KODAMA101}) in relation to the Hamiltonian constraint $H(x)$,
are also true:
\begin{eqnarray}
\label{KODAMA10}
e^{-ia\widehat{Q}}\widehat{V}e^{ia\widehat{Q}}=e^{-\lambda{a}}\widehat{V};~~
e^{-ib\widehat{V}}\widehat{Q}e^{ib\widehat{V}}=\widehat{Q}+\lambda{b}\widehat{V}.
\end{eqnarray}
\noindent
\par \noindent (iii) Step 3: Next, we introduce an
involution operation $*$ on $B_{aux}$, which defines an algebra
$B_{aux}^{*}$.  So one must have have $(A^{*})^{*}=A$,
$(B^{*})^{*}=B$, and $(AB)^{*}=A^{*}B^{*}$ and similarly for the
products in opposite order.  The quantum analogues of $A$ and $B$
must be self-adjoint operators such that
\begin{eqnarray}
\label{KODAMA9}
A^{*}\longrightarrow\widehat{A}^{\dagger};~~B^{*}\longrightarrow\widehat{B}^{\dagger};~~(\widehat{A}\widehat{B})^{\dagger}=\widehat{B}^{\dagger}\widehat{A}^{\dagger}
\end{eqnarray}
\noindent for all $A,B$.  Applying this to (\ref{KODAMA8}) one sees
that
\begin{eqnarray}
\label{IMPOSE1}
[\widehat{Q}^{\dagger},\widehat{V}^{\dagger}(x)]=-i\lambda\widehat{V}^{\dagger}(x);~~[\widehat{Q}^{\dagger},\widehat{Q}^{\dagger}]=[\widehat{V}^{\dagger}(x),\widehat{V}^{\dagger}(y)]=0.
\end{eqnarray}
\noindent The result is that the set of elementary observables $S$
is not only closed under Poisson brackets and complex conjugation,
but also is consistent with the involution operation $*$.  The
existence of self-adjoint operators $\widehat{Q}$ and $\widehat{V}$
requires that a physical Hilbert space ${\bf H}_{Phys}$ be defined,
such that the $\dagger$ operation acts by Hermitian conjugation.  We
use the term `physical' since all of the constraints will have
inherently been implemented.\par \noindent (iv) Step 4: Ultimately
we will construct a linear $*$- representation $\pi$ of the abstract
algebra $B^{(*)}_{aux}$ via linear operators on the physical Hilbert
space ${\bf H}_{Phys}$.  For the remaining steps of the quantization
procedure we will proceed along a different path to the one
presented in \cite{ASHTEKAR}, as we will now be addressing the
quantization of the Hamiltonian constraint and the physical states
from a different interpretation.
\par \indent It could be argued that $S$ is not `large enough' since it
essentially consists merely of three elements.  After all, there are
an infinite number of sufficiently regular functions on the original
Ashtekar phase space $(A^a_i,\widetilde{E}^i_a)$ which cannot be
obtained as a sum of products of elements of $S$.  On the other
hand, $S$ is `large enough' to admit nontrivial, unitary
representations of its associated Lie algebra on a natural Hilbert
space in which all physical states satisfy, and arise from, the
affine algebra.

\subsection{Construction of the Hilbert space ${\bf H}$}

Having put in place the necessary elements, we will now proceed with
the construction of the Hilbert space ${\bf H}$.  Let us define for
our fiducial vector $\vert\eta\rangle=\vert{0},0\rangle$, a
gauge-invariant, diffeomorphism-invariant state lying in the kernel
of the quantum constraints $\vert{0},0\rangle\in{Ker}\widehat{C}$,
where
$\widehat{C}=\{\widehat{G}_a(x),\widehat{H}_i(x),\widehat{H}(x)\}$
are the Gauss' law, diffeomorphism and Hamiltonian constraints.  By
$Ker\{\widehat{H}(x)\}$, we mean that the Hamiltonian constraint
acting on $\vert\eta\rangle$ is given by the affine commutation
relation
\begin{eqnarray}
\label{KODAMA11}
\lambda\widehat{V}(x)\vert{0},0\rangle=[i\widehat{Q},\widehat{V}(x)]\vert{0},0\rangle,
\end{eqnarray}
\noindent which is a restatement of $\widehat{H}(x)=0$.  In this
interpretation, $\vert{0},0\rangle$ plays the role of a ``seed" from
which other physical states can be obtained. A generic basis (
denoted by  $\{|\alpha\rangle\}$ ) of a representation space of the
affine group has resolution of unity, $\int \mu {d\alpha}
|\alpha\rangle\langle\alpha| =I$, with respect to an appropriate
measure $\mu$ (see, for instance the discussion in Appendix A). The
fiducial state can be expanded as $\vert{0},0\rangle = \int d\alpha
C(\alpha)|\alpha\rangle$ for some coefficients $C(\alpha)$. Since
$Q$ is a hermitian generator, $e^{-ia\widehat{Q}}$,  where $a$ is an
arbitrary real-valued dimensionless numerical constant, is a unitary
affine group element whose representation is
$D^Q_{\alpha'\alpha}=\langle
\alpha'|e^{-ia\widehat{Q}}|\alpha\rangle$. It follows that
\begin{eqnarray}
\label{KODAMA112}
\Bigl[i\widehat{Q},\widehat{V}(x)\Bigr]{e}^{-ia\widehat{Q}}\vert{0},0\rangle\nonumber\\
=\int\mu {d\alpha'}
\Bigl[i\widehat{Q},\widehat{V}(x)\Bigr]|\alpha'\rangle\int
\mu{d\alpha}  \langle\alpha'|{e}^{-ia\widehat{Q}}
|\alpha\rangle\langle\alpha\vert{0},0\rangle
\nonumber\\
=\int\mu{d\alpha'} \Bigl[i\widehat{Q},\widehat{V}(x)\Bigr]|\alpha'\rangle\int \mu d\alpha D^Q_{\alpha'\alpha}\langle\alpha\vert{0},0\rangle\nonumber\\
=\int \mu {d\alpha'} \lambda\widehat{V}(x)|\alpha'\rangle\int \mu{d\alpha}  \langle\alpha'|{e}^{-ia\widehat{Q}} |\alpha\rangle\langle\alpha\vert{0},0\rangle\nonumber\\
=\lambda\widehat{V}(x){e}^{-ia\widehat{Q}} \vert{0},0\rangle;
\end{eqnarray}
where we have used the fact that an affine representation
$\{|\alpha\rangle\}$  satisfies, by definition,
$\Bigl[i\widehat{Q},\widehat{V}(x)\Bigr]|\alpha'\rangle
=\lambda\widehat{V}(x)|\alpha'\rangle \quad \forall\, \alpha'$. But
(\ref{KODAMA112}) demonstrates that $\vert{a},0\rangle=
{e}^{-ia\widehat{Q}} \vert{0},0\rangle$ which is
diffeomorphism-invariant since $[H_i, Q]=0$ also satisfies
\begin{eqnarray}
\label{KODAMA15}
\lambda\widehat{V}(x)\vert{a},0\rangle=[\widehat{V}(x),
-i\widehat{Q}]\vert{a},0\rangle,
\end{eqnarray}
which is none other than the vanishing of the local Hamiltonian
constraint $\widehat{H}(x)$ acting on the dilated state
$\vert{a},0\rangle$. So given that
$\vert{0},0\rangle\in{Ker}\widehat{C}$, it follows that
$\vert{a},0\rangle\in{Ker}\widehat{C}$ as well i.e.
$\vert{a},0\rangle \in {\bf H}_{Phys}$.

Using the unitary group element $e^{-ibV(x)}$ at each spatial point,
and noting that $[V(x), V(y)]$ commutes $\forall \,x,y \in \Sigma$,
the global volume $V=\lim_{\Delta x \rightarrow 0}\sum_{x \in
\Sigma}V(x)\Delta x$, corresponds to the unitary group element
$e^{-ib\widehat{V}}=\lim_{\Delta x \rightarrow 0}\Pi_{x\in \Sigma}
e^{-ib{\widehat V(x)}\Delta x}$ , where $b$ is an arbitrary
real-valued numerical constant of mass dimension $[b]=3$. This can
be used to construct the diffeomorphism-invariant state, translated
in the carrier space of the affine group,
$\vert{0},b\rangle=e^{-ib\widehat{V}}\vert{0},0\rangle.$ Similar
considerations will then yield
\begin{eqnarray}
\label{KODAMA12}
\Bigl[i\widehat{Q},\widehat{V}(x)\Bigr]{e}^{-ib\widehat{V}}\vert{0},0\rangle\nonumber\\
=\int\mu {d\alpha'}
\Bigl[i\widehat{Q},\widehat{V}(x)\Bigr]|\alpha'\rangle\int
\mu{d\alpha}  \langle\alpha'|{e}^{-ib\widehat{V}}
|\alpha\rangle\langle\alpha\vert{0},0\rangle
\nonumber\\
=\int\mu{d\alpha'} \Bigl[i\widehat{Q},\widehat{V}(x)\Bigr]|\alpha'\rangle\int \mu d\alpha D^V_{\alpha'\alpha}\langle\alpha\vert{0},0\rangle\nonumber\\
=\int \mu {d\alpha'} \lambda\widehat{V}(x)|\alpha'\rangle\int \mu{d\alpha}  \langle\alpha'|{e}^{-ib\widehat{V}} |\alpha\rangle\langle\alpha\vert{0},0\rangle\nonumber\\
=\lambda\widehat{V}(x){e}^{-ib\widehat{V}} \vert{0},0\rangle;
\end{eqnarray}
which leaves us with
\begin{eqnarray}
\label{KODAMA19}
\lambda\widehat{V}(x)\vert{0},b\rangle=[\widehat{V}(x),
-i\widehat{Q}]\vert{0},b\rangle,
\end{eqnarray}
\noindent which again expresses the vanishing of the local
Hamiltonian constraint $\hat{H}(x)$ acting on the translated state
$\vert{0},b\rangle$.  As $[H_i,V]=0$, given that
$\vert{0},0\rangle\in{Ker}\widehat{C}$, then
$\vert{0},b\rangle\in{Ker}\widehat{C}$ as well.\par \indent

\subsection{General quantum Affine group element}

Having illustrated the idea for physical states corresponding to
transformations by $a$ and by $b$ individually, we will now consider
the two transformations applied together.  Define the general affine
coherent state
\begin{eqnarray}
\label{KODAMA20}
\vert{a},b\rangle=U(a,b)\vert{0},0\rangle={e}^{-ia\widehat{Q}}e^{-ib\widehat{V}}\vert{0},0\rangle.
\end{eqnarray}
By using the resolution of unity, and invoking the linearity of the
action of the group elements $D^Q_{\alpha'\alpha}$ and
$D^V_{\alpha'\alpha}$, it follows that
\begin{eqnarray}
\label{KODAMA24}
\lambda\widehat{V}(x)\vert{a},b\rangle=i[\widehat{Q},\widehat{V}(x)]\vert{a},b\rangle
\end{eqnarray}
also holds  for the coherent state $\vert{a},b\rangle$.  The result
is that given a fiducial state
$\vert{0},0\rangle\in{Ker}{\widehat{C}}$, it follows that an
arbitrary diffeomorphism-invariant coherent state
$\vert{a},b\rangle=U(a,b)\vert\eta\rangle\in {\bf H}_{Phys} $ lies
in the kernel of all of the constraints.

The coherent nature with respect to the uncertainty in $Q$ and the
total volume $V$,  and other properties, of these states are
discussed in Appendix A. In particular, the uncertainty relation
{\it which depends on the cosmological constant}  is $\langle\Delta
\widehat{V}\rangle^{2}\langle\Delta
\widehat{Q}\rangle^{2}\geq\frac{\lambda^2}{4}{\langle
\widehat{V}\rangle^{2}}$.

\subsection{Normalizability and inner product}

If a fiducial state $\vert{0},0\rangle$ satisfies the admissibility
condition (\ref{ADMISSIBILITY}), then all states $\vert{a},b\rangle$
are unitarily related to $\vert{0},0\rangle$, and form an
overcomplete basis of physical coherent states.\footnote{These
results and the results which follow are independent of the specific
carrier representation space of the affine group.}  Let the fiducial
state be normalized such that
\begin{eqnarray}
\label{KODAMA25} \langle0,0\vert{0},0\rangle=1.
\end{eqnarray}
\noindent We would like to find the inner product between two states
$\vert{a},b>$ and $\vert{a}^{\prime},b^{\prime}>$.  This is given by
\begin{eqnarray}
\label{KODAMA26} \langle{a}^{\prime},b^{\prime}\vert{a},b\rangle
=\langle0,0\vert{e}^{ib^{\prime}\widehat{V}}e^{ia^{\prime}\widehat{Q}}e^{-ia\widehat{Q}}e^{-ib\widehat{V}}\vert{0},0\rangle\nonumber\\
=\langle0,0\vert{e}^{ib^{\prime}\widehat{V}}e^{i(a^{\prime}-a)\widehat{Q}}e^{-ib\widehat{V}}\vert{0},0\rangle.
\end{eqnarray}
\noindent Proceeding from (\ref{KODAMA26}) and using the trick
$I=e^{-\widehat{A}}e^{\widehat{A}}$, we have
\begin{eqnarray}
\label{KODAMA27} \langle{a}^{\prime},b^{\prime}\vert{a},b\rangle
=\langle0,0\vert{e}^{i(a^{\prime}-a)\widehat{Q}}e^{-i(a^{\prime}-a)\widehat{Q}}e^{ib^{\prime}\widehat{V}}
e^{i(a^{\prime}-a)\widehat{Q}}e^{-ib\widehat{V}}\vert{0},0\rangle\nonumber\\
=\langle0,0\vert{e}^{i(a^{\prime}-a)\widehat{Q}}\hbox{exp}\Bigl[ib^{\prime}e^{-\lambda(a^{\prime}-a)}\widehat{V}\Bigr]e^{-ib\widehat{V}}\vert{0},0\rangle\nonumber\\
\langle0,0\vert{e}^{i(a^{\prime}-a)\widehat{Q}}\hbox{exp}\Bigl[i\bigl(-b+b^{\prime}e^{-\lambda(a^{\prime}-a)}\bigr)\widehat{V}\Bigr]\vert{0},0\rangle\nonumber\\
=\langle0,0\vert{a}-a^{\prime},b-b^{\prime}e^{\lambda(a-a^{\prime})}\rangle.
\end{eqnarray}
\noindent As (\ref{KODAMA27}) shows, the overlap between two states
is equivalent to performing two affine group transformations in
opposite directions, and the coherent states will thus have unit
norm.

\section{Summary and discussion}

The affine quantization program in conjunction with the relevant
steps of the algebraic quantization procedure yield a natural
Hilbert space and facilitate the construction of gravitational
coherent physical states ${\vert a, b\rangle}$ and other physical
states which satisfy all the constraints of 4-dimensional Lorentzian
GR.

  While we have performed a quantization of a Poisson-closed algebra consisting essentially of three elements $Q,V(x)$ and $I$ (the identity operator), it is important to note that these objects do not separate the points of the full classical phase space of GR.  Nevertheless, they are of utmost importance to the Hamiltonian constraint, and through them an affine representation admitting natural Hilbert space structures consistent with the full Lorentzian theory can be defined. The total volume $V$ , $I$, and $Q$  operators as diffeomorphism and gauge invariant entities form a very limited subset of invariant phase space elements.
However, the group theoretical aspect of the affine quantization
program enables the infusion of new results and techniques from
wavelet transform theory into quantum gravity, wherein the powerful
coherent state machinery becomes available.  In this approach one
can dispense with direct reference to the quantum operators
$\widehat{V}(x)$, $\widehat{V}$ and $\widehat{Q}$, and think of the
carrier space in terms of coherent states encoding the group action.
The results of this paper depend only on the existence of fiducial
vectors satisfying the admissibility condition, and are independent
of the specific representation or polarization of the carrier space.
Within the context of this work, {\it all} physical states of
quantum gravity with cosmological constant {\it must} come from
representations of the affine algebra, since it is the full {\it
local} Hamiltonian constraint, and not just a mini-superspace
version, that has been reformulated as an affine algebra. We have
demonstrated that the affine quantization program of Klauder can
indeed be applicable to 4-dimensional full Lorentzian signature
quantum gravity.  It is also intriguing that the formulation of the
Hamiltonian constraint as an affine algebra is predicated upon a
non-vanishing cosmological constant (current observational bounds
place $\Lambda L^2_{Planck} \sim 10^{-120}$); and a fundamental
uncertainty relation of the form $\frac{\langle\Delta
\widehat{V}\rangle}{\langle
\widehat{V}\rangle}\langle\Delta\widehat{Q}\rangle\geq\frac{\lambda}{2}=
2\pi \Lambda L^2_{Planck}$ may govern all physical states of quantum
gravity.

\section*{Appendix A. Measure, inner product, and coherent states
associated with the affine group} The action of the affine group
$G_{aff}$ on the real numbers has the following matrix
representation
\begin{eqnarray}
U(a,b)\equiv \left(\begin{array}{cc}
a & b\\
0 & 1\\
\end{array}\right)
,
\end{eqnarray}
\noindent which has a natural action on itself by left-matrix
multiplication via the group multiplication law
\begin{eqnarray}
\label{ESSS5} U(a,b)U(a',b')=U(aa',b+ab').
\end{eqnarray}
\noindent The affine group has invariant left Haar and right Haar
measures $d\mu_l$ and $d\mu_r$ respectively, given by
\begin{eqnarray}
\label{HAAR} d\mu_l(a,b)={{{da}\wedge{db}} \over
{a^2}};~~d\mu_r(a,b)={{{da}\wedge{db}} \over a}.
\end{eqnarray}
\noindent While these two measures are equivalent in the sense of
measure theory, they are not the same measure.  Hence the affine
group is not a unimodular group.  With respect to the unitary
representation of the group one sees, from exponentiation of
(\ref{AFFINERELATION}), that the parametrization of the general
group element as the linear operator
\begin{eqnarray}
\label{ESSS4} U(a,b)=e^{-iD\hbox{ln}a}e^{-ibq}
\end{eqnarray}
\noindent correctly reproduces the group multiplication law
(\ref{ESSS5}).\footnote{Note that group composition for the unitary
form (\ref{ESSS4}) must be performed in the opposite order as the
matrix version (\ref{ESSS5}).}  Using any normalized fiducial vector
$\vert\eta\rangle$ with $\langle\eta\vert\eta\rangle=1$, one can
define an overcomplete basis of unit vectors as
\begin{eqnarray}
\label{ESSS6} \vert{a},b\rangle=U(a,b)\vert\eta\rangle.
\end{eqnarray}
\noindent So for each fiducial vector $\vert\eta\rangle$, there
exists a set of states labeled by $a$ and $b$.  Since there is a
large class of possible fiducial vectors $\vert\eta\rangle$, then
one has certain freedom in the choice of Hilbert spaces (one could
go further, utilizing the admissibility condition for fiducial
vectors in order to construct reproducing kernel Hilbert spaces (See
e.g. \cite{KLAUDER2})).\par \indent Equation (\ref{ESSS6}) bears an
analogy to continuous wavelet transform theory \cite{ALI}, where
$\vert{a},b\rangle$, a set of coherent states, plays the role of the
wavelet transform of a mother wavelet (signal) $\vert\eta\rangle$. A
necessary condition is that $\vert\eta\rangle$ satisfy a certain
admissibility condition predicated on its existence as an element of
the set ${\psi}\in{L}^2(R,dz)$.  The admissibility condition is
\begin{eqnarray}
\label{ADMISSIBILITY}
c_{\psi}=2\pi\int^{\infty}_{-\infty}\frac{d\xi}{\vert\xi\vert}\vert\phi(\xi)\vert^2<\infty,
\end{eqnarray}
\noindent where $\phi$ is the Fourier transform of ${\psi}$.  Note
that the unitary action of (\ref{ESSS4}) on $\vert\eta\rangle$
translates in the language of functions into
\begin{eqnarray}
\label{FUNCTIONS}
{\psi}'(z)=U(a,b){\psi}=\vert{a}\vert^{-1/2}{\psi}\Bigl({{z-b} \over
a}\Bigr),
\end{eqnarray}
\noindent where $b\in{R}$ and $a\neq{0}$.  That $U(a,b)$ is unitary
is evident in the fact that it preserves the Hilbert space norm
\begin{eqnarray}
\label{NORM}
\Vert{\psi}\Vert^2=\int_{-\infty}^{\infty}dz\vert{\psi}'(z)\vert^2.
\end{eqnarray}
\noindent In the context of quantum gravity, the square
integrability of the representation $U(a,b)$ implies the existence
of fiducial vectors ${\psi}$ for which the matrix element
$\langle{U}(a,b){\psi}\vert{\psi}\rangle$ is square integrable as a
function of the labels $a$ and $b$ with respect to the left Haar
measure
\begin{eqnarray}
\label{HAAR1}
\int_{G_{aff}}d\mu_l(a,b)\vert\langle{U}(a,b){\psi}\vert{\psi}\rangle\vert^2<\infty.
\end{eqnarray}
\noindent Additionally, since the Fourier transform is a linear
isometry, it follows that the Fourier transformed version of
(\ref{ESSS4}) also provides a unitary representation of the affine
group in its action on $\phi$.  Both representations are
irreducible.

%\subsection{Coherent states associated with the affine algebra}
In this paper we will use the `affine conjugate' pair
$(\widehat{V},\widehat{Q})$\,\, where $V$ is the integral of $V(x)$
over all space (the volume operator), to describe quantum gravity.
Our application is somewhat different from that described by the
affine conjugate pair $(\widehat{q}_{ij},\widehat{\pi}^{k}_{l})$ due
to Klauder, namely the spatial metric and the field
$\widehat{\pi}^{k}_{l}$\,\,related to the conjugate momentum
$\widehat{\pi}^{kl}$ of the spatial metric. Closure of the
constraints algebra in canonical quantum gravity described by
$(\widehat{q}_{ij},\widehat{\pi}^{kl})$\,\, produces second-class
constraints \cite{KLAUDER3}.  This is unlike the case as described
in the Ashtekar variables $(A^a_i, \widetilde{E}^i_a)$, where the
constraints algebra is first-class (at least at the classical
level). Hence for this paper, we will use the affine commutator of
$\widehat{Q}$\,\,and $\widehat{V}$.
%, which at the same time addresses the issues mentioned regarding LQG.\par

In the following, we want to construct the coherent state framework
according to the article \cite{KLAUDER2} by using the `affine
conjugate' pair $(\widehat{Q},\widehat{V})$, as another application
for the affine representation.  We also can shadow the treatment of
the spatial metric matrix field $g$\,\, and the momentum matrix
field $\pi$\,\, to describe this topic.  Articles \cite{KLAUDER1},
\cite{KLAUDER3} provide further descriptions.  Here, the commutation
relation for $(\widehat{V},\widehat{Q})$\,\, is suitable for the
one-dimension affine algebra, $[\widehat{V}, -i\widehat{Q}]=
\lambda\widehat{V}$.  This algebra follows from integration over all
spatial points of the local fundamental commutation relation
$[\widehat{V}(x), -i\widehat{Q}]= \lambda\widehat{V}(x)$, which is
the local Hamiltonian constraint $H(x)=0$.

One can write the following realization of $Q$ (for ease of
discussion here we absorb $1/\lambda$ in the definition of $Q$)
\begin{eqnarray}
\label{A1}
\widehat{Q}=\frac{1}{2}[\widehat{\theta}\widehat{V}+\widehat{V}\widehat{\theta}]\qquad,\qquad\widehat{\theta}=-\imath\frac{\partial}{\partial\widehat{V}},
\end{eqnarray}
in terms of $V$ which is also equal to
\begin{eqnarray}
\label{A2}
\widehat{Q}=\frac{1}{2}[-\imath\frac{\partial}{\partial\widehat{V}}\widehat{V}+\widehat{V}(-\imath\frac{\partial}{\partial\widehat{V}})]
=-\frac{\imath}{2}-\imath\widehat{V}\frac{\partial}{\partial\widehat{V}}.
\end{eqnarray}
With respect to a polarization on $V-space$ one has the inner
product
\begin{eqnarray*}
\langle\phi|\psi\rangle=\int_{0}^{\infty}\phi(\widehat{V})^{\ast}\psi(\widehat{V})d\widehat{V},
\end{eqnarray*}
consistent with the positivity of $V$ as required by the affine
algebra.  The operator $\widehat{Q}$\,\,generates the unitary
dilations in the representation space:
\begin{eqnarray*}
e^{-\imath
a\widehat{Q}}\psi(\widehat{V})=e^{-\frac{a}{2}}\psi(e^{-a}\widehat{V})
\end{eqnarray*}
\begin{eqnarray*}
\|e^{-\imath a\widehat{Q}}|\psi\rangle\|=\||\psi\rangle\|.
\end{eqnarray*}
From the definition for unitary operators $U(a,b)=e^{-\imath
a\widehat{Q}}e^{-\imath b\widehat{V}}$\,\, subject to the
composition rule $U(a',b')U(a,b)=U(a'+a,b+e^a)$, one can construct a
set of coherent states:
\begin{eqnarray*}
|a,b\rangle=U(a,b)|\eta\rangle
\end{eqnarray*}
where $|\eta\rangle$\,\,is an unspecified normalized fiducial vector
in the representation space. The family of coherent states provides
a resolution of unity in the form:
\begin{eqnarray*}
N^{-1}\int_{-\infty}^{\infty}db\int_{-\infty}^{\infty}da
e^{a}|a,b\rangle\langle a,b|= I,
\end{eqnarray*}
where $N=2\pi\int_{0}^{\infty}d\widehat{V}
\widehat{V}^{-1}|\eta(\widehat{V})|^{2}<\infty$\,\,, is finite.\par
\indent For minimum uncertainty states, one takes the Heisenberg
uncertainty principle into consideration. In general quantum
mechanics, any two observable operators $\widehat{A}$\,\,,
$\widehat{B}$\,\, are consistent with the uncertainty relation:
\begin{eqnarray*}
\langle(\Delta \widehat{A})^{2}\rangle\langle(\Delta
\widehat{B})^{2}\rangle\geq\frac{1}{4}|\langle[\widehat{A},\widehat{B}]\rangle|^{2}
\end{eqnarray*}
But on the affine representation space, any two self-adjoint
operators can be consistent with the general affine uncertainty
relation: (here, the two operators are $\widehat{V}$\,\,and
$\widehat{Q}$):
\begin{eqnarray*}
\langle\Delta \widehat{V}\rangle^{2}\langle\Delta
\widehat{Q}\rangle^{2}\geq\lambda^{2}\frac{\langle
\widehat{V}\rangle^{2}}{4}
\end{eqnarray*}
We can use the general analysis just like that in \cite{KLAUDER2} to
prove the affine uncertainty relation by taking a normalized
fiducial vector with of form
$\eta_{1}(\widehat{V})=C_{1}(\alpha',\beta')\widehat{V}^{\alpha'}e^{-\beta'\widehat{V}}$.
Where $\alpha',\beta'$\,\,are positive real coefficients,
$C_{1}(\alpha',\beta')$\,\,is the normalization constant.

\section{Appendix B. Proof that the real part of the Chern--Simons functional Poisson-commutes with the volume element}
 Consider the
Chern-Simons functional:
$I_{CS}[A]=\frac{1}{2}\int_{\Sigma}[A^{a}\wedge
dA_{a}+\frac{1}{3}\epsilon^{abc}A_{a}\wedge A_{b}\wedge A_{c}]$.
Using the definition (\ref{THECONNECTION}), and defining the one
form $k^{a}=K^{a}_idx^i$, expansion of the Chern-Simons functional
in $k$ and $\Gamma$ leads straightforwardly to the expression
\begin{eqnarray*}
 I_{CS}[A]=I_{CS}[\Gamma]+\gamma\int_{M}R^{\Gamma}_{a}\wedge k^{a}
+\frac{\gamma^{2}}{2!}\int_{M}k^{a}\wedge
(D^{\Gamma}k)_{a}+\frac{\gamma^{3}}{3!}\int_{M}\epsilon^{abc}k_{a}\wedge
k_{b}\wedge k_{c},
\end{eqnarray*}
where
$R^{\Gamma}_{a}=d\Gamma_{a}+\frac{1}{2}\epsilon_{a}^{bc}\Gamma_{b}\wedge\Gamma_{c}$
is the curvature two form of the connection one form $\Gamma^a$.
Consider the Poisson bracket of the volume functional and the
Chern-Simons functional (for brevity we suppress the label of
spatial points $x$. The results which follow equally apply to the
global volume functional $V$, just as they apply to the local
$V(x)$.  Note also that
$\{F,\sqrt{\hbox{det}\widetilde{E}}\}=\frac{1}{2\sqrt{\hbox{det}\widetilde{E}}}\{F,\hbox{det}\widetilde{E}\}$
for all $F$). To wit,

\begin{eqnarray*}
\{{V},I_{CS}[A]\}&=&\{{V},I_{CS}[\Gamma]\}+\gamma\int_{M}\{{V},R^{\Gamma}_{a}\wedge
k^{a}\} +\frac{\gamma^{2}}{2!}\int_{M}\{{V},k^{a}\wedge
(D^{\Gamma}k)_{a}\}
\end{eqnarray*}
\begin{eqnarray*}
+\frac{\gamma^{3}}{3!}\int_{M}\{{V},\epsilon^{abc}k_{a}\wedge
k_{b}\wedge k_{c}\}
\end{eqnarray*}
\begin{eqnarray*}
 &=&\{{V},I_{CS}[\Gamma]+\frac{\gamma^{2}}{2!}\int_{M}k^{a}\wedge
(D^{\Gamma}k)_{a}\}+\{{V},\gamma\int_{M}R^{\Gamma}_{a}\wedge
k^{a}+\frac{\gamma^{3}}{3!}\int_{M}\epsilon^{abc}k_{a}\wedge
k_{b}\wedge k_{c}\}
\end{eqnarray*}
\begin{eqnarray*}
 &=&\{{V},Re[(I_{CS}[A])\}+\{{V},Im(I_{CS}[A])\} \quad (\gamma= \pm
i)
\end{eqnarray*}
\begin{eqnarray*}
 &=&\{{V},Im[I_{CS}[A]]\}.
\end{eqnarray*}
The result can be explained by the following observations:\par
\noindent (1)The term $\{{V},I_{CS}[\Gamma]\}$ is zero because
$\Gamma$\,\,is a function only of $\widetilde{E}$, thus
Poisson-commuting with $V$.
\\
(2)The term $\frac{\gamma^{2}}{2!}\int_{M}\{{V},k^{a}\wedge
(D^{\Gamma}k)_{a}\}$ also vanishes. We note that
\begin{eqnarray*}
\frac{\gamma^{2}}{2!}\int_{M}\{{V},k^{a}\wedge
(D^{\Gamma}k)_{a}\}&\propto&
\{\epsilon_{lmn}\epsilon_{dbc}\widetilde{E}^{ld}\widetilde{E}^{mb}\widetilde{E}^{nc},\int\epsilon^{ijk'}k_{k'a}D^{\Gamma}_{i}k_{j}^{a}\}
\end{eqnarray*}
\begin{eqnarray*}
&=&\int
\lmn\epsilon_{dbc}\epsilon^{ijk'}\{\widetilde{E}^{ld}\widetilde{E}^{mb}\widetilde{E}^{nc},k_{k'a}D^{\Gamma}_{i}k^{a}_{j}\}
\end{eqnarray*}
\begin{eqnarray*}
&=&\int \lmn\epsilon_{dbc}\epsilon^{ijk'}\bigl(
\{\widetilde{E}^{ld},k_{k'a}D^{\Gamma}_{i}k^{a}_{j}\}\widetilde{E}^{mb}\widetilde{E}^{nc}
+\widetilde{E}^{ld}\{\widetilde{E}^{mb},k_{k'a}D^{\Gamma}_{i}k^{a}_{j}\}\widetilde{E}^{nc}
\end{eqnarray*}
\begin{eqnarray}
\label{a3}
&&+\widetilde{E}^{ld}\widetilde{E}^{mb}\{\widetilde{E}^{nc},k_{k'a}D^{\Gamma}_{i}k^{a}_{j}\}
\bigr).
\end{eqnarray}
We now focus on the Poisson bracket
$\{\widetilde{E}^{ld},k_{k'a}D^{\Gamma}_{i}k^{a}_{j}\}$, namely
\begin{eqnarray*}
\{\widetilde{E}^{ld},k_{k'a}D^{\Gamma}_{i}k^{a}_{j}\}=\{\widetilde{E}^{ld},k_{k'a}\}D^{\Gamma}_{i}k^{a}_{j}
+k_{k'a}\{\widetilde{E}^{ld},D^{\Gamma}_{i}k^{a}_{j}\}.
\end{eqnarray*}
Note that
\begin{eqnarray*}\{\widetilde{E}^{ld},D^{\Gamma}_{i}k^{a}_{j}\}
=D^{\Gamma}_{i}\{\widetilde{E}^{ld},k^{a}_{j}\}\propto
\delta^{l}_{j}\delta^{d}_{a} (D^{\Gamma}_{i}\delta^{3})
\end{eqnarray*}
\noindent where we have used the shorthand notation
$\delta^3\equiv\delta^{(3)}(x,y)$.  From this Poisson bracket
$\{\widetilde{E}^{ld},D^{\Gamma}_{i}k^{a}_{j}\}=\widetilde{E}^{ld}(D^{\Gamma}_{i}k^{a}_{j})
-(D^{\Gamma}_{i}k^{a}_{j})\widetilde{E}^{ld}\propto
\delta^{l}_{j}\delta^{d}_{a} (D^{\Gamma}_{i}\delta^{3})$,
$\Longrightarrow
(D^{\Gamma}_{i}k^{a}_{j})\widetilde{E}^{ld}=\widetilde{E}^{ld}(D^{\Gamma}_{i}k^{a}_{j})-\delta^{l}_{j}\delta^{d}_{a}
(D^{\Gamma}_{i}\delta^{3})$. So we have
\begin{eqnarray*}
\{\widetilde{E}^{ld},k_{k'a}D^{\Gamma}_{i}k^{a}_{j}\}=\{\widetilde{E}^{ld},k_{k'a}\}D^{\Gamma}_{i}k^{a}_{j}
+k_{k'a}\{\widetilde{E}^{ld},D^{\Gamma}_{i}k^{a}_{j}\}
\\
=\{\widetilde{E}^{ld},k_{k'a}\}D^{\Gamma}_{i}k^{a}_{j}+k_{k'a}D^{\Gamma}_{i}\{\widetilde{E}^{ld},k^{a}_{j}\}
\propto\delta^{l}_{k'}\delta^{d}_{a}
\delta^{3}(D^{\Gamma}_{i}k^{a}_{j})+k_{k'a}\delta^{l}_{j}\delta^{d}_{a}
(D^{\Gamma}_{i}\delta^{3})
\end{eqnarray*}
The three terms in equation (\ref{a3}) will each yield the same
result, which can be seen by a relabeling of indices.  So it
suffices to illustrate the calculation for one term.  In this
process, we will also use $\{\widetilde{E}^{ia},k_{jb}\}\propto
\delta^{i}_{j}\delta^{a}_{b}\delta^{3}$.\\
Consider the term $\int \lmn\epsilon_{dbc}\epsilon^{ijk'}
\{\widetilde{E}^{ld},k_{k'a}D^{\Gamma}_{i}k^{a}_{j}\}\widetilde{E}^{mb}\widetilde{E}^{nc}
$\,\,in equation (\ref{a3}):
\begin{eqnarray*}
 \int \lmn\epsilon_{dbc}\epsilon^{ijk'}
\{\widetilde{E}^{ld},k_{k'a}D^{\Gamma}_{i}k^{a}_{j}\}\widetilde{E}^{mb}\widetilde{E}^{nc}\\
 \propto \int
\lmn\epsilon_{dbc}\epsilon^{ijk'}\delta^{l}_{k'}\delta^{d}_{a}
\delta^{3}(D^{\Gamma}_{i}k^{a}_{j})\widetilde{E}^{mb}\widetilde{E}^{nc}
+\int
\lmn\epsilon_{dbc}\epsilon^{ijk'}k_{k'a}\delta^{l}_{j}\delta^{d}_{a}
(D^{\Gamma}_{i}\delta^{3})\widetilde{E}^{mb}\widetilde{E}^{nc}\\
\propto\int\epsilon^{lij}\delta^{3}(D^{\Gamma}_{i}k_{ja}){\widetilde{E}}^a_l+\int(D^{\Gamma}_{i}\delta^{3})\epsilon^{ijk'}k_{k'a}\widetilde{E}^{a}_{j}
\end{eqnarray*}
\begin{eqnarray*}
=-\int\delta^{3}D^{\Gamma}_{i}(\epsilon^{ilj}\widetilde{E}^{a}_{l}k_{ja})+\int(D^{\Gamma}_{i}\delta^{3})\epsilon^{ijk'}\widetilde{E}^{a}_{j}k_{k'a}
\end{eqnarray*}
\begin{eqnarray*}
\propto
-\int\delta^{3}D^{\Gamma}_{i}[\epsilon_{da'b'}\widetilde{E}^{ja'}\widetilde{E}^{ib'}k_{j}^{d}]+\int(D^{\Gamma}_{i}\delta^{3})[\epsilon_{da'b'}\widetilde{E}^{k'a'}\widetilde{E}^{ib'}k_{k'}^{d}]
\end{eqnarray*}
\begin{eqnarray*}
=\int\delta^{3}D^{\Gamma}_{i}[\widetilde{E}^{ia}(\epsilon_{a'da}\widetilde{E}^{ja'}k_{j}^{d})]-\int(D^{\Gamma}_{i}\delta^{3})[\widetilde{E}^{ia}(\epsilon_{a'da}\widetilde{E}^{k'a'}k_{k'}^{d})]
\end{eqnarray*}
\begin{eqnarray*}
=\int\delta^{3}D^{\Gamma}_{i}[\widetilde{E}^{ia}G_{a}]-\int(D^{\Gamma}_{i}\delta^{3})[\widetilde{E}^{ia}G_{a}]
=2\int\delta^{3}D^{\Gamma}_{i}[\widetilde{E}^{ia}G_{a}]=0
\end{eqnarray*}
\\
wherein $G_{a}=\epsilon_{aa'd}\widetilde{E}^{ja'}k_{j}^{d}$\,\,is
just the Gauss Law constraint in the ADM triad formulation of
gravity. So equation (\ref{a3}) provides no contribution and $
\frac{\gamma^{2}}{2!}\int_{M}\{\widehat{V},k^{a}\wedge
(D^{\Gamma}k)_{a}\}=0
$ as desired.\\
(3)Therefore, the result is that $\{\widehat{V},Im(I_{CS}[A])\}
=\{\widehat{V},I_{CS}[A]\}$ since $\{\widehat{V},Re(I_{CS}[A])\}
=0$.

\section{Acknowledgements}
This work has been supported in part by the Office of Naval Research
under Grant No. N-000-1412-WX-30191, the National Science Council of
Taiwan under Grant No. NSC 101-2112-M-006 -007 -MY3, and the
National Center for Theoretical Sciences, Taiwan.

\end{document}